\documentclass[fleqn,usenatbib]{mnras}

\usepackage{newtxtext,newtxmath}

\usepackage[T1]{fontenc}
\usepackage{ae,aecompl}

\usepackage{graphicx}	
\usepackage{amsmath}	

\usepackage{float,tikz,bbm,tabularx}
\usepackage[ruled,vlined ]{algorithm2e}
\usepackage{lineno}
\usepackage[toc,page]{appendix}
\usepackage{verbatim}
\usepackage{hyperref}

\usepackage{soul}

\DeclareMathOperator*{\argmin}{arg\,min}
\DeclareMathOperator{\sign}{sgn}

\newcommand\norm[1]{\left\lVert#1\right\rVert}

\def\aap{Astronomy \& Astrophysics}
\def\apj{The Astrophysical Journal}
\def\mnras{Monthly Notices of the Royal Astronomical Society}
\def\pasp{{PASP}}               
\def\procspie{\ref@jnl{Proc.~SPIE}}   

\title[TLDR: Time Lag/Delay Reconstructor]{TLDR: Time Lag/Delay Reconstructor}

\author[M.D. Anderson]{M.D. Anderson$^{1,2}$\thanks{E-mail: manderson@astro.gsu.edu}, F. Baron$^{1,2}$\thanks{E-mail: baron@astro.gsu.edu} and M.C. Bentz$^{1}$\thanks{E-mail: bentz@astro.gsu.edu}\\
$^{1}$Georgia State University, 25 Park Place NE, Atlanta GA 30303-2911, U.S.A\\
$^{2}$Center for High Angular Resolution Astronomy, P. O. Box 5060, Atlanta, GA 30302-5060, U.S.A}

\date{Accepted XXX. Received YYY; in original form ZZZ}

\pubyear{2021}

\begin{document}
\label{firstpage}
\pagerange{\pageref{firstpage}--\pageref{lastpage}}
\maketitle

\begin{abstract}
We present the Time Lag/Delay Reconstructor (TLDR), an algorithm for reconstructing velocity delay maps in the Maximum A Posteriori framework for reverberation mapping. Reverberation mapping is a tomographical method for studying the kinematics and geometry of the broad-line region of active galactic nuclei at high spatial resolution. Leveraging modern image reconstruction techniques, including Total Variation and Compressed Sensing, TLDR applies multiple regularization schemes to reconstruct velocity delay maps using the Alternating Direction Method of Multipliers. Along with the detailed description of the TLDR algorithm we present test reconstructions from TLDR applied to synthetic reverberation mapping spectra as well as a preliminary reconstruction of the $H\beta$ feature of Arp 151 from the 2008 Lick Active Galactic Nuclei Monitoring Project.

\end{abstract}

\begin{keywords}
galaxies:active,nuclei -- techniques:image processing
\end{keywords}



\section{Introduction}\label{sec:Intro}

In an active galactic nucleus (AGN) clouds of photoionized gas near the Supermassive Black Hole (SMBH) at the galaxy's center emit highly broadened emission lines due to the extreme kinematics of the region. The area in which these broad lines are emitted is known as the broad-line region (BLR). Rapid variation of the broad line emission in response to variations in the continuum emission has long been observed for both radio quiet and radio loud AGN \citep{1967ApJ...150L.177S,1970ApJ...159L.147C,1974MNRAS.169..357P,1988PASP..100...18P,1997ARA&A..35..445U}.

While the precise mechanism behind the observed variability is unknown, a number of viable theories exist, including, but not limited to, simple accretion onto the SMBH, magnetic field interactions within the accretion disk, or some combination of processes varying in contribution over various timescales \citep{2006ASPC..360..265C}. Regardless of the mechanism at play, the release of energy is observed as continuum light which propagates outward from the central region of the AGN. During the process, clouds of gas are ionized at different times depending on their distance from the accretion disk. As the photoionized gas recombines, it re-emits the light as an emission line feature subject to the kinematics of the cloud. This is to say that any observed change above the noise in continuum emission from the AGN will be followed by a change in emission lines in the BLR separated by a delay corresponding to the difference in path length between the directly observed continuum emission, the resulting emission line feature, and the observer. 
\begin{figure}
	\begin{center}
	\includegraphics[width=210px]{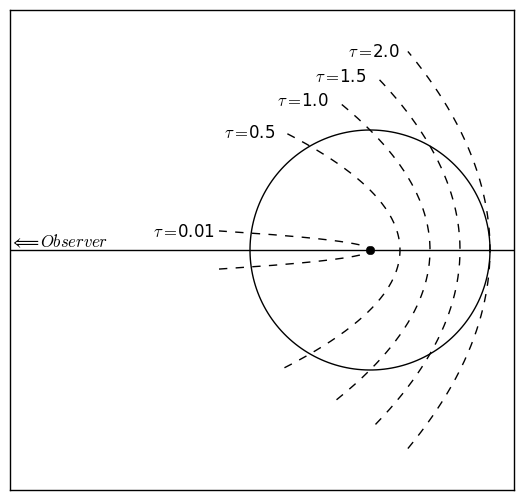}

 \caption{Classic isodelay surface schematic of an AGN showing how the various delays sample the entire area of the AGN consisting of a thin ring of spectral emitting material about the central SMBH. Based on Figure 1 from \protect\cite{1993PASP..105..247P}, with permission.}\label{fig:isodelay}
\end{center}
\end{figure}

The classic example is that of a thin ring as described in \cite{1993PASP..105..247P}, shown in schematic form in Fig.~\ref{fig:isodelay}, where the observer lies in plane of the figure at a sufficient distance along the horizontal line that light traveling from the thin ring to the observer arrive in effectively flat wavefronts.The thin ring of gas at a radius of one light-day from the central black hole is ionized by a pulse of light described by a $\delta$-function. The ionizing pulse reaches the gas that comprises the shell at the same time. The gas is ionized and the energy is then re-emitted by the gas, subject to Doppler effects due to the kinematics of the region. Some of this re-emitted light is along the line of sight to the observer. Light that is re-emitted at the far side of the ring must travel back across the region, a distance of twice the ring's radius, such that it arrives at the observer with a delay of $2r/c$. Following the arc of the ring from the near side to the far side, the observer will see the re-emitted light arrive with delay times covering the range from 0 to $2r/c$ with a bulk average delay of $r/c$. In Fig.~\ref{fig:isodelay} the dashed lines show lines of constant delay, known as isodelay surfaces) across the thin ring at varying delay times $\tau$.

Moving from the thin ring to a more realistic distribution of gas around the SMBH results in more complex patterns in the observed response. Reverberation mapping~(RM) is the tomographical method used to study the arrangement of photoionized gas near accreting black holes \citep{1982ApJ...255..419B}. At present, RM is primarily used to measure the masses of the SMBH at the heart of AGNs \citep{2000ApJ...540L..13P,2009ApJ...705..199B,2015ApJ...806..128D,2015ApJS..216....4S,1538-3873-127-947-67}. RM results also establish the radius-luminosity relationship \citep{2009hst..prop11662B,2013ApJ...767..149B}, which allows us to estimate SMBH masses across all redshifts and consequently consider their influence on the evolution of galaxies across cosmological time \citep{2009MNRAS.398...53B}.

Studying the kinematics of AGNs at the geometric scales provided by RM represents a substantial impetus behind RM campaigns \citep{1993PASP..105..247P,2004PASP..116..465H}. By providing complementary information on scales smaller than may be probed via Integral Field Spectroscopy \citep{Storchi_Bergmann_2019} or Interferometry \citep{2020A&A...635A..92G}, RM has the potential to bridge the gap in spatial scale between inflows and outflows and the central engine of AGNs. 

One goal of RM in the inverse problem approach is to recover the velocity delay map (VDM), of an observed AGN. The VDM is a representation of the BLR where the geometry is encoded in the observed delay time and line of sight velocity of the re-emitting gas. Mathematical recovery of the VDM results from solving an ill-posed inverse problem, meaning that the number of degrees of freedom to reconstruct (the VDM) is greater than the effective number of data points (flux-varying spectra). A conventional solution to this problem is regularized maximum likelihood or Maximum A Posteriori  (MAP) in a Bayesian framework. Most software packages for RM rely on this formalism \citep{1994ASPC...69...23H,1994ASPC...69...53K,2015MNRAS.454..144S}, where the $\chi^{2}$ of the reconstructed map is minimized along a set of regularizers that enforce the prior expectations on the model, the exception being a direct modeling approach \citep{2014MNRAS.445.3055P}.

Classic examples of regularizers are positivity and maximum entropy (to enforce smoothness). Note that without these priors, the reconstruction algorithm would over-fit the $\chi^{2}$, giving rise to spurious artifacts which would make the VDM un-interpretable. On the other hand, regularizers must remain generic, i.e. as non-committal as possible, so as not to bias the reconstruction towards a non-physical solution. As shown in other fields of study such as interferometry \citep{schutz:hal-01123500,2010ISPM...27...97T}, and medical imaging \citep{CTRecon,Cao:07,Yao:15}, it is the quality of the regularization and its strength that determine the actual mapping fidelity of a given piece of software. Unfortunately, currently available software packages for RM offer little to no flexibility in terms of regularizers, and are mostly centered on smoothing, which \textit{de facto} limit the interpretation of the data. From recent advances in the theory of Compressed Sensing, we now know that the key to optimally regularize such inverse problems is tapping into the sparsity of the reconstructed signals \citep{4472240}. In image reconstruction, sparsity is the idea that a signal may be represented by a number of non-zero coefficients that is small relative to the total number of coefficients in the signal \citep{davenport2011introduction,doi:10.1137/110837681}. RM is a poster example for Compressed Sensing, in that the VDMs that have been previously recovered for various AGNs are sparse.

The Time Lag/Delay Reconstruction algorithm (TLDR) presented here utilizes the Alternating Direction Method of Multipliers (ADMM), as well as two-dimensional regularization across the temporal and spectral axes of RM data, to reconstruct VDMs for RM.

This paper serves as an introduction to image reconstruction in the MAP framework as applied to RM. It will discuss the basics of reverberation mapping in Section~\ref{Sec:RM}, the inverse problem approach and appropriate regularizers in Section~\ref{Sec:InvProbl}, the basics of ADMM in Section~\ref{Sec:ADMM}, the construction of the TLDR algorithm as an application of ADMM in Section~\ref{Sec:TLDR}, basic reconstruction tests in Section~\ref{Sec:Testing}, and offer a preliminary reconstruction of the Arp 151 $H\beta$ feature in Section~\ref{Sec:TestingArp151}.  

\section{Reverberation Mapping}\label{Sec:RM}

Reverberation mapping is a tomographical imaging process in which rapid changes in continuum emission from the center of an AGN is used to map the kinematics and geometry of the gas that re-emits the continuum emission as line emission. While the precise relationship between the ionizing and emitted radiation is unknown, a number of assumptions serve as the backbone of reverberation mapping. These assumptions as laid down in \cite{1993PASP..105..247P} are:

\begin{enumerate}
  \item The continuum emission originates from a compact isotropically emitting central source.
  \item The light-travel time between the central source and the clouds of the BLR is the primary source of delay.
  \item A simple relationship between the ionizing emission of the central source and the response observed in BLR exists. This relationship need not be strictly linear in nature.
\end{enumerate}
In reality, continuum emission does not appear in $\delta$-function pulses, but rather as continuous variations. Both the spectral line and continuum emission must be treated as continuous functions. Doing so, the observed line emission, $\mathbf{L}(t,\lambda)$ and continuum emission, $\mathbf{C}(t)$, are functions of time which are related by the VDM $\bf{x}$ \citep{1994ASPC...69....1P}. Mathematically, the emission line response is the result of the continuum being convolved with the VDM, also called the transfer function \citep{1982ApJ...255..419B},

\begin{equation}\label{eqn:trueconvolution}
\mathbf{L}\left(t,\lambda\right)=\int_{0}^{\infty}\mathbf{x}\left(\tau,\lambda\right)\mathbf{C}\left(t-\tau\right)d\tau.
\end{equation}

\noindent Due to the discrete nature of observations, the convolution is not continuous, but discrete which appears as:

\begin{equation}\label{eqn:truediscconv}
\mathbf{L}[t,\lambda] = \sum_{\tau=0}^N\mathbf{x}[\tau,\lambda]\mathbf{C}[t-\tau].
\end{equation}

\noindent The nature of the element-wise convolution allows the construction of a mapping matrix $\bf{H}$ comprised of elements from  the continuum so that equation~\ref{eqn:truediscconv} can be represented purely as the linear relation

\begin{equation}\label{eqn:convolution}
\mathbf{L} =  \mathbf{H}\mathbf{x}.
\end{equation}

\noindent The matrix $\bf{H}$ is constructed with rows corresponding to the number of observations in the spectra taken at time $t$ and columns corresponding to the delay times $\tau$ such that each element in the matrix is a value of the continuum flux measured at time $t-\tau$. The resulting mapping matrix $\mathbf{H}$ is a Toeplitz matrix representing the 1-dimensional convolution. Since the entire problem is linear, recovering the VDM is a straightforward inverse problem. 

It is evident in Eqn.~\ref{eqn:convolution} that this method requires that the continuum data must exist at the proper sample times $t-\tau$ for all spectral sample dates $t$ and all delay times $\tau$. The required high cadence represents a serious challenge for all RM observational campaigns. In earlier work on RM, the requirement that continuum observations exist on a grid has been met by linearly interpolating the continuum data \citep{1994ASPC...69...23H,1994ASPC...69...53K}. Recently, it has become common to model the continuum data in some way where both Damped Random Walk \citep{2011ApJ...735...80Z,2009ApJ...698..895K} and Gaussian Process \citep{2011ApJ...730..139P} models have been used to interpolate continuum observations.

\section{Inverse Problem Approach}\label{Sec:InvProbl}

In order to find the VDM, we adopt here the general framework for MAP. Prior knowledge about the system, such as positivity, smoothness, etc., is introduced in the form of regularization terms \citep{2005opas.conf..397T} that are minimized alongside the neg-log-likelihood of the data. The VDM solution $\mathbf{x}_\text{opt}$ is given by:
\begin{equation}
\mathbf{x}_\text{opt} = \argmin_{\mathbf{x} \in \mathbb{R}^{M\times N}}\left\{f_{data}\left(\mathbf{x}\right)+f_{prior}\left(\mathbf{x}\right)\right\} \label{eqn:COP}
\end{equation}
Assuming Gaussian statistics based on flux levels, exposure times, and the Central Limit Theorem, the $f_{data}$ term is commonly known as the $\chi^{2}$ of the model relative to the data. In the case that Gaussian statistics do not hold and Poisson noise dominates, ADMM could still be used. Here, construction of the $\chi^2$ involves combining the discrete convolution of equation~\ref{eqn:convolution} with knowledge of the error in the data,

\begin{equation}\label{eq:likelihood}
f_{data}(\mathbf{x}) = \frac{1}{2}\chi^{2}(\mathbf{x}) = \frac{1}{2} \norm{\mathbf{L}-\mathbf{Hx}}^{2}_{W}=\frac{1}{2} \left(\mathbf{Hx}-\mathbf{L}\right)^\top
\mathbf{W}\left(\mathbf{Hx}-\mathbf{L}\right),
\end{equation}
\noindent where $\mathbf{W}$ is the inverse covariance matrix of the emission line data $\mathbf{L}$.

The second term in equation~\ref{eqn:COP}, is the regularization term where the prior knowledge and expectations of the problem are enforced. The regularization term consists of the weighted sum of all regularizers, so that $f_\mathrm{prior}$ can be written:
\begin{equation}\label{eqn:regform}
f_{prior}(\mathbf{x})= \sum_{i}\mathit{\mu_{i}}R_{i}(\mathbf{x}).
\end{equation}

\noindent The combination of each regularizer, $R_{\mathrm{i}}$, and its associated regularization hyper-parameter, $\mu_{\mathrm{i}}$, allows each regularization term to be weighted individually allowing for flexibility in the reconstruction process. These functions serve to penalize the likelihood term in the minimization equation (equation~\ref{eqn:COP}). The effect of these penalties is to increase the value of the function within the minimizer, making these solutions less probable in the Bayesian framework. Selection of regularizers for a reconstruction is an important step in setting up the reconstruction. Aside from being important to the eventual solution the use of different regularization strategies may impact the problem to be fed into the minimization algorithm. 

For the RM problem, there are a number of regularizations that are natural choices. Foremost among these is positivity. Due to the nature of the AGN system creating the continuum pulses and the linear relationship between the pulses and the re-emission from the reverberating medium, only positive delays have any physical meaning. Negative values can only arise in the reconstruction where the linear model does not hold. The positivity regularizer works as a filter where negative values are simply projected to zero.

Since previously reconstructed VDMs are visibly sparse in the spectral axis \citep{2010ApJ...720L..46B,2013ApJ...764...47G}, another obvious regularizer choice is that of the $\ell_{\mathrm{1}}$-norm. More specifically, in the spectrum of an AGN at wavelengths where no broad emission lines are present, the VDM is expected to be zero for all delay times. As previously mentioned, this sparsity makes the RM problem a particularly good example for compressed sensing. As such, the so called $\ell_{\mathrm{1}}$-norm is well suited for reconstructing VDMs. The $\ell_{\mathrm{1}}$-norm favors the fewest number of nonzero terms in $\bf{x}$. In the case of RM this means that solutions with more zero-valued VDMs will be favored over any other solutions, matching with the expected sparsity in the VDM. Regularization by the $\ell_1$-norm is accomplished according to the Soft Thresholding scheme of \cite{donoho1995noising} with an additional enforcement of positivity.

As an alternative to the $\ell_\mathrm{1}$-norm, the square of the $\ell_\mathrm{2}$-norm favors images with many zero valued coefficients. It is however, slightly different mathematically and its primary purpose is to promote smoothness in the image. The $\ell_\mathrm{2}$-norm penalizes larger pixel values more than small ones due to the square of the values seen in equation. Reducing the value of the $\ell_\mathrm{2}$-norm-squared drives the total amount of flux in the image down and brings the maximum and minimum flux values closer together, promoting smoother images.

Perhaps most importantly, the VDM is expected to be smooth both along the temporal axis and the spectral axis. The temporal axis is expected to be smooth due to the relationship between absorbed and emitted light from the gas clouds being mapped in the BLR. The spectral axis will be smooth due to the kinematics of the region and the Doppler broadening of the emission lines themselves. Compressed sensing theory prescribes to enforce smoothness based on sparsity, either that of the gradient of the solution, or even more generally that of its wavelet transform. In both cases, regularization along the different axes is better treated as two individual regularization steps, allowing the regularization along each axis to be weighted differently.

In the case of gradient sparsity, the regularizer is known as anisotropic total variation (TV) and given by the $\ell_\mathrm{1}$-norm of the gradient of the image \citep{1992PhyD...60..259R}. Minimizing this regularization term favors $\bf{x}$ such that the sum over the gradient is as small as possible, which is to say the solutions with the least variations are favored. TV in this case is constructed using a finite difference matrix operator to represent the gradient, and the TV regularization term can be easily calculated and minimized using the same Soft Thresholding scheme as the standard $\ell_1$-norm, albeit without the enforcement of positivity. 



\section{ADMM}\label{Sec:ADMM}
In order to solve the problem posed by equation~\ref{eqn:RMminProb}, ADMM is used. ADMM was developed as a combination of the method of multipliers and the dual ascent method \citep{Boyd:2011:DOS:2185815.2185816,eckstein2012augmented}. The convergence behavior of ADMM is well documented in general \citep{Wang2018,Nishihara:2015:GAC:3045118.3045156} and for linear and quadratic problems specifically \citep{doi:10.1137/120878951,doi:10.1137/120886753}. This section is comprised of a discussion of ADMM for Global Variable Consensus Optimization (GVCO) and how it solves problems of the form of equation~\ref{eqn:RMminProb}.

GVCO ADMM seeks to find the  solution to the general GVCO problem,

\begin{equation}\label{eqn:GVCOprob}
\begin{aligned}
& \argmin_{\mathbf{x_i}}
& & \sum_{i=1}^{N}f_{i}\left(\mathbf{x}_{i}\right) \\
& \text{subject to}
& & \mathbf{x}_{i}-\mathbf{z}=0,  i=1,...,N.
\end{aligned}
\end{equation}

\noindent While having a different form than the general equation for solving an inverse problem in the MAP framework (equation~\ref{eqn:COP}), equation~\ref{eqn:GVCOprob} represents the same problem. The differences lie in the construction of the ADMM algorithm. Key in the construction of ADMM is a variable splitting. In ADMM, the problem must be separable so that the general function can be split into two groups of functions, $f_\mathrm{i}(\mathbf{x}_\mathrm{i})$ and $g(\mathbf{z})$. ADMM will then seek to bring these two function into consensus so that at convergence $\mathbf{x}_\mathrm{i}=\mathbf{z}$.

The ADMM algorithm is an augmented Lagrangian method, so the algorithm relies on the augmented Lagrangian in its minimization steps. The augmented Lagrangian takes the general form,

\begin{equation}\label{eqn:GVCORAL}
\begin{aligned}
\mathcal{L} \left( \right. & \left. \mathbf{x}_\mathrm{1}, \ldots,\mathbf{x}_{\mathrm{N}},\mathbf{z},\mathbf{y} \right)=  \\
	 & \sum_{i=1}^{N}\left(f_i\left(\mathbf{x}_i\right)+\mathbf{y}_{i}^{\top}\left(\mathbf{x}_{i}-\mathbf{z}\right)+\frac{\rho_{i}}{2}\norm{\mathbf{x}_{i}-\mathbf{z}}^2_{2}\right)
\end{aligned}
\end{equation}
 where $\mathbf{y}$ represents the Lagrange multipliers and $\rho$ the penalty parameter. 
 ADMM then proceeds iterating and following these three steps, where the current iteration number is denoted by a superscript $\mathrm{k}$:
\begin{enumerate}
\item \label{it:minx} Minimize each term, subscript $\mathrm{i}$, of the augmented Lagrangian with respect to the corresponding $\mathbf{x}_\mathrm{i}$,
\begin{equation}\label{eq:admm1C}
\mathbf{x}_{i}^{k+1}=\argmin_{\mathbf{x}_{i}} \left(f_{i}\left(\mathbf{x}^{k}_{i}\right)+\mathbf{y}_{i}^{\top,{k}}\left(\mathbf{x}^{k}_{i}-\mathbf{z}^{k}\right)+\frac{\rho_{i}}{2}\norm{\mathbf{x}^{k}_{i}-\mathbf{z}^{k}}_{2}^2 \right).
\end{equation}
\item Minimize the augmented Lagrangian with respect to $\bf{z}$,
\begin{equation}\label{eq:admm2C}
\mathbf{z}^{{k+1}}=\argmin_{\mathbf{z}}\left( \sum_{{i=1}}^{{N}} -\mathbf{y}_{i}^{\top,{k}}\mathbf{z}^{{k}}+\frac{\rho_{i}}{2}\norm{\mathbf{x}_{i}^{{k+1}}-\mathbf{z}^{k}}^{2}_{2}\right).
\end{equation}
\item \label{it:multup} Update the Lagrange Multipliers,
\begin{equation}\label{eq:admm3C}
\mathbf{y}_{i}^{{k+1}}=\mathbf{y}_{i}^{k}+\rho_{i}\left(\mathbf{x}_{i}^{{k+1}}-\mathbf{z}^{{k+1}}\right).
\end{equation}
\end{enumerate}
\noindent This procedure repeats until whatever convergence parameters are in place are satisfied. This method offers flexibility in that numerous regularization terms can be added with relatively little difficulty. The method also offers the potential for great computational speed in that many steps and subroutines in the algorithm can be implemented in parallel.

\section{TLDR}\label{Sec:TLDR}
 Inspired by recent implementations of ADMM in interferometry \citep{2014JOSAA..31.2334S,Ferrari2015,10.1117/12.926862}, the TLDR algorithm is built as an implementation of GVCO ADMM constructed to solve the RM problem given by equation~\ref{eqn:RMminProb}. Through the assignment of new variable names, one for each regularization term used, equations~\ref{eq:admm1C},~\ref{eq:admm2C}, and~\ref{eq:admm3C} can be expanded to build the algorithm. Within TLDR, the $\chi^2$ and $\ell_2$-norm smoothing regularizations operate on $\bf{X}$, the positivity regularization operates on $\bf{P}$, the compressed sensing regularizer operates on $\bf{N}$, and the total variation reguarizer operates on $\bf{T}$.  The new variables and their associated penalty parameters and Lagrange multipliers help to distinguish mathematical operations in the code, and are listed in Table~\ref{tab:Variables}. For convenience, the $z$ variable is assigned to $\bf{Z}$.
 
\begin{table*}
\begin{minipage}{\textwidth}

\caption{ADMM variable assignments within TLDR.}
\label{tab:Variables}
 \begin{tabular}{c c c c c c} 
 
\hline\hline 
Regularizer & Original Term & Variable Assignment & New Term & Penalty Parameter & Lagrange Multiplier  \\ [0.5ex] 
\hline 
Smoothing & $\norm{\bf{Hx-L}}^2_2+\frac{\mu_{\ell_2}}{2}\norm{\bf{x}}_2^2$ & $\bf{x}_\mathrm{0}=\mathbf{X}$ & $\norm{\bf{HX-L}}^2_F+\frac{\mu_{\ell_2}}{2}\norm{\bf{X}}_2^2$ &$\rho_\mathrm{X}$ & $\mathbf{Y_\mathrm{X}}$ \\
Positivity & $\mathbbm{1}_{\mathbb{R}^{+}}(\mathbf{x})$ & $\bf{x_\mathrm{1}}=\mathbf{P}$ & $ \mathbbm{1}_{\mathbb{R}^{+}}(\mathbf{P})$ &$\rho_\mathrm{P}$ & $\mathbf{Y_\mathrm{P}}$ \\ 
$\ell_\mathrm{1}$-norm & $\mu_{\ell_\mathrm{1}}\norm{\mathbf{x}}_\mathrm{1}$ & $\mathbf{x}_\mathrm{2}=\mathbf{N}$ & $\mu_{\ell_\mathrm{1}}\norm{\mathbf{N}}_\mathrm{1}$&$\rho_\mathrm{N}$ & $\mathbf{Y}_\mathrm{N}$ \\ 
Total Variation & $\mu_{\mathrm{T}}\norm{\mathbf{\nabla}\mathbf{x}}_\mathrm{1}$ & $\mathbf{x}_\mathrm{3}=\mathbf{\nabla}\mathbf{Z}=\mathbf{T}$ &$\mu_{\mathrm{T}}\norm{\mathbf{T}}_\mathrm{1}$& $\rho_\mathrm{T}$ & $\mathbf{Y}_\mathrm{T}$ \\[1ex] 
\hline 
\end{tabular}

\end{minipage}
\end{table*}

\noindent With the new variables in place, the problem can be rewritten as follows,

\begin{equation}\label{eqn:RMminProb}
\begin{aligned}
 &	 \argmin_{\mathbf{X}}	& \frac{1}{2}\left(\mathbf{HX} - \mathbf{L}\right)^{\top}\mathbf{W}\left(\mathbf{HX}-\mathbf{L}\right)+\frac{\mu_{X}}{2}\norm{\mathbf{X}}_{2}^2+\\
 & &\mu_{N}\norm{\mathbf{N}}_1 +\mathbbm{1}_{\mathbb{R}^{+}}\left(\mathbf{P}\right) +\mu_{{T}}\norm{\mathbf{T}}_{{1}} \\
& \text{subject to} & \mathbf{X},\mathbf{P},\mathbf{N}=\mathbf{Z}; \mathbf{T}=\nabla\mathbf{Z}
\end{aligned}.
\end{equation}

\noindent Expanding equation~\ref{eqn:GVCORAL} using the new variables and the problem as written in equation~\ref{eqn:RMminProb} allows us to easily write the full Augmented Lagrangian.

\begin{equation}\label{eqn:FullAugLagrangian}
\begin{aligned}
\mathcal{L} = &\left(\mathbf{HX}-\mathbf{L}\right)^\top\mathbf{W}\left(\mathbf{HX}-\mathbf{L}\right) +\frac{\mu_{\ell_{2}}}{2}\norm{\mathbf{X}}_2^2  \\
 +&\mathbf{Y_X}^{\top}\left(\mathbf{X}-\mathbf{Z}\right)+\frac{\rho_X}{2}\norm{\mathbf{X}-\mathbf{Z}}^2_2 \\
  +& \mathbbm{1}_{\mathbb{R}^{+}}\left(\mathbf{P}\right) 
 +\mathbf{Y_P}^{\top}\left(\mathbf{P}-\mathbf{Z}\right)+\frac{\rho_P}{2}\norm{\mathbf{P}-\mathbf{Z}}^2_2 \\
+&\mu_{\ell_1}\norm{\mathbf{N}}_{1}  
 +\mathbf{Y_N}^{\top}\left(\mathbf{N}-\mathbf{Z}\right)+\frac{\rho_N}{2}\norm{\mathbf{N}-\mathbf{Z}}^2_2\\
+& \mu_{T}\norm{\mathbf{T}}_{1}
 +\mathbf{Y_{T}}^{\top}\left(\mathbf{T}-\mathbf{\nabla}\mathbf{Z}\right)+\frac{\rho_T}{2}\norm{\mathbf{T}-\mathbf{\nabla}\mathbf{Z}}^2_2
\end{aligned}
\end{equation}

\noindent TLDR proceeds through the steps of ADMM outlined in Section~\ref{Sec:ADMM} shown as pseudocode in algorithm~\ref{alg:1}.

\begin{algorithm}\label{alg:1}
\SetAlgoLined

 initialization\: $\mathbf{X},\mathbf{P}, \mathbf{N} = {\mathbf Z}$; $\mathbf{T}=\mathbf{\nabla}\mathbf{Z}$\\
 calculate Initial VDM \\
 \While{ converged == False}{
  minimization of Augmented Lagrangian w.r.t. $\mathbf{X}$ \\
  minimization of Augmented Lagrangian w.r.t. regularization terms $\mathbf{P}$, $\mathbf{N}$, $\mathbf{T}$\\
  minimization of Augmented Lagrangian w.r.t. $\mathbf{Z}$	 \\
  update Lagrange Multipliers $\mathbf{Y}_\mathrm{X}$, $\mathbf{Y}_\mathrm{P}$, $\mathbf{Y}_\mathrm{N}$, {$\mathbf{Y}_\mathrm{T}$} \\
  check for convergence \\
  }
 \caption{Pseudocode for TLDR's ADMM block.}
\end{algorithm}

Aside from the Lagrange multipliers update, which is performed exactly as equation~\ref{eq:admm3C}, the steps in algorithm~\ref{alg:1} are non-trivial and each warrants some discussion. So each will be discussed in order.

\subsection{Initial VDM}\label{SubSecInitialVDM}
The TLDR algorithm could begin from any initial VDM, but the optimal starting delay map should be as close to the final solution as possible. This reduces the amount of parameter space over which the pixel values in the delay map must move to find the best values. By starting near to the final solution, the number of iterations required by the algorithm is reduced thereby reducing the amount of time required for the algorithm to find the final solution. However, care must be taken as it is possible to introduce artifacts into the initial VDM that will not easily be removed within the rest of the algorithm.

By isolating the first term in the augmented Lagrangian, a singly-regularized least squares equation for the Reverberation Mapping problem is found:

\begin{equation}
f\left(\mathbf{X}\right)=\left(\mathbf{HX}-\mathbf{L}\right)^\top\mathbf{W}\left(\mathbf{HX}-\mathbf{L}\right) + \frac{\mu_{\ell_2}}{2}\norm{\mathbf{X}}_2^2.
\end{equation}

\noindent Minimizing, by taking the derivative of the function $f(\mathbf{X})$, setting it equal to zero, and solving for $\bf{X}$, yields the initial VDM:

\begin{equation}\label{eqn:TikInit}
\mathbf{X}^0= \left(\mathbf{H}^{\top}\mathbf{WH}+\mu_{\ell_{2}}\mathbf{I}\right)^{-1}\left(\mathbf{H}^{\top}\mathbf{WL}\right).
\end{equation}

\noindent This method for analytically solving the $\ell_\mathrm{2}$-squared regularized least squares problem  is known as Tikhonov Regularization \citep{Tikhonov:1963} or Ridge Regression \citep{doi:10.1080/00401706.1970.10488634}.

\subsection{Minimization with respect to~X}\label{subsec:minwrtx}
The minimization for $\bf{X}$ can be found analytically by solving the derivative of equation~\ref{eqn:FullAugLagrangian}, set  equal to zero, for $\bf{X}$, yielding:

\begin{equation}\label{eq:minwrtXB}
\begin{aligned}
\mathbf{X}^{k+1}=&\left(\mathbf{H}^{\top}\mathbf{W}\mathbf{H}+\left(\mu_X+\rho_X\right)\mathbf{I}\right)^{-1}\left(\mathbf{H}^\top\mathbf{W}\mathbf{L}+\rho_X\mathbf{Z}-\mathbf{Y}_X^k\right).
\end{aligned}
\end{equation}

Due to the spectral dependence of $\mathbf{W}$, the minimization with respect to $\mathbf{X}$ must be carried out spectral line by spectral line so that equation~\ref{eq:minwrtXB} is used to find $\mathbf{X}$ by finding $\mathbf{X}_{\lambda}$ using each unique $\mathbf{W}_{\lambda}$, however, this step can be parallelized and the inversion step is constant. The pseudocode for solving equation~\ref{eq:minwrtXB} is shown in algorithm~\ref{alg:min_wrt_x}.

\begin{algorithm}\label{alg:min_wrt_x}
\SetAlgoLined
 \For{ $\lambda$}{
  $\mathbf{X}_\lambda = \left( H^\top W_\lambda H + \left(\mu_X+\rho_X\right)I\right)^{-1}$
  $\left(\mathbf{H}^\top W_\lambda L_\lambda +\rho_XZ_\lambda - Y_{X,\lambda}\right)$ 
  }
 \caption{Pseudocode for TLDR's minimization of the Augmented Lagrangian with respect to $\bf{X}$.}
\end{algorithm}

\subsection{Minimization w.r.t. Regularizers} \label{subsec:RegMins}
The minimization of the augmented Lagrangian for each of the regularized variables involving the $\ell_1$-norms of $\mathbf{N}$ and $\mathbf{T}$ is identical, though the input changes. The minimization problem can be written using a stand-in variable $\mathbf{r}_\mathrm{i}$,
\begin{equation}\label{eqn:RegMAP}
   \argmin_{\mathbf{r_i}}  f\left(\mathbf{r}_{i}\right) +\frac{\rho_{\mathbf{r}_{i}}}{2}\norm{\mathbf{r}_{i}-\mathbf{\tilde{r}}_{i}}^2_{2}.
\end{equation}
when posing $\mathbf{\tilde{r}}_\mathrm{i}=\mathbf{Z}-\mathbf{Y}_{\mathbf{r}_\mathrm{i}}/\rho_{\mathbf{r}_\mathrm{i}}$. This is as a MAP problem, solvable by proximal operators. A proximal operator allows the minimization to be replaced with analytic functions that can be evaluated quickly within the ADMM loop, e.g. the proximal operator for the $\ell_1$-norm is known as soft-thresholding \citep{donoho1995noising}. The solution to the minimization is the projection 

\begin{equation}\label{eqn:STproxop}
 \mathbf{r_i^+} =\sign\left(\tilde{r}_i\right)  \max\left(\lvert\tilde{r}_i\rvert-\frac{\mu_r}{\rho_r},0\right).
\end{equation}

\noindent Though the projection is identical, the input for the projected solution for $\bf{T}$ must include the gradient. So that for $\bf{T}$, $\mathbf{\tilde{r}}_\mathrm{T}=\mathbf{\nabla Z}-\mathbf{Y}_{\mathbf{T}}/\rho_{\mathbf{T}}$. In practice, the projection relating to negative values $\bf{\tilde{r}_i}\leq -\mu_r / \rho_r$ in the VDM are rejected as an additional enforcement of positivity.

Similarly, for the positivity regularizer, equation~\ref{eqn:RegMAP} holds and only the projection changes. Where the projection is simply,

\begin{equation}\label{eqn:posproxop}
\mathbf{P}^+ = \max\left(0,\mathbf{Z} - \frac{\mathbf{Y_P}}{\rho_P}\right).
\end{equation}

In effect, the proximal operator for positivity simply rejects values where $\bf{Z} <\frac{\bf{Y_P}}{\rho_P}$. 

\subsection{Minimization w.r.t. Z}\label{subsec:zMin}
To find the analytical solution for $\bf{Z}$, the same procedure as used for the minimization of $\mathbf{X}$ is used. Begin by taking the derivative of equation~\ref{eqn:FullAugLagrangian} with respect to Z, set equal to zero and solve for $\bf{Z}$. Yielding,

 \begin{equation}\label{eqn:zsolution}
 \begin{aligned}
 \mathbf{Z}^{k+1} = &\left( \left(\rho_X + \rho_P + \rho_N\right)\mathbf{I}+\rho_{T}\nabla^\top \nabla\right)^{-1}\\ 
& \left(\mathbf{Y}_X+\rho_X\mathbf{X}+\mathbf{Y}_P+\rho_P\mathbf{P}+\mathbf{Y}_N+\rho_N\mathbf{N}+\nabla^\top\mathbf{Y}_{T}+\rho_{T}\nabla^\top\mathbf{T}\right).
 \end{aligned}
 \end{equation}

\noindent The minimization w.r.t. $\bf{Z}$ benefits from the lack of spectral dependence in the solution. In its implemented form, $\bf{Z}$ is calculated very quickly as matrix multiplication with Fast Fourier Transforms to handle the gradients. Implementing the the forward difference gradient operators as an implementation of the convolution theorem allows the Z update step to be calculated by only multiplication. Note also, that the inversion used is a constant, meaning that the Z update step is extremely fast.

\subsection{Convergence Testing}
In TLDR, convergence testing is carried out as outlined in \cite{Boyd:2011:DOS:2185815.2185816} $\S 3.3.1$. At each iteration $k$, we define the primal residual $\mathbf{r}^k$, dual residual $\mathbf{s}^k$, and their tolerances $\epsilon^{pri}$ and $\epsilon^{dual}$ as:

\begin{equation}
\mathbf{r}^k = \left(\mathbf{X}^k-\mathbf{Z}^k , \mathbf{N}^k-\mathbf{Z}^k , \mathbf{P}^k-\mathbf{Z}^k ,\frac{1}{2}\left( \mathbf{T}^k-\nabla\mathbf{Z}^k\right)\right)
\end{equation}

\begin{equation}
\begin{aligned}
\mathbf{s}^k = &\left(\rho_X \left(\mathbf{Z}^k - \mathbf{Z}^{k-1}\right), \rho_N \left(\mathbf{Z}^k - \mathbf{Z}^{k-1}\right),\right.\\ &\left.\rho_P\left(\mathbf{Z}^k - \mathbf{Z}^{k-1}\right) , \rho_{T}\left( \mathbf{Z}^k - \mathbf{Z}^{k-1}\right)\right)
\end{aligned}
\end{equation}

\begin{equation}
\begin{aligned}
    \epsilon^{pri} = \epsilon^{abs} + \epsilon^{rel} & \max \left\{  \norm{\mathbf{X}^k}_2,\norm{\mathbf{N}^k}_2,\norm{\mathbf{P}^k}_2,\right.\\
    & \left.\frac{1}{2}\norm{\mathbf{T}^k}_2,\norm{\mathbf{Z}^k}_2\right\},
\end{aligned}
\end{equation}
\noindent and
\begin{equation}
\begin{aligned}
    \epsilon^{dual} = \epsilon^{abs}  + \epsilon^{rel}  & \max \left\{ \norm{\mathbf{Y_X}^k}_2,\norm{\mathbf{Y_N}^k}_2,\norm{\mathbf{Y_P}^k}_2, \right.\\
  & \left.\frac{1}{2}\norm{\nabla^\top\mathbf{Y_{T}}^k}_2,\right\},
\end{aligned}
\end{equation}
\noindent where $\epsilon^{pri}$ and $\epsilon^{dual}$ are the global tolerances on the primal and dual residuals respectively.

TLDR is said to have converged when both the norm of the primal residual and norm of the dual residual have dropped below their respective tolerances $\epsilon^{pri}$ and $\epsilon^{dual}$. Which is to say, TLDR has converged when $\norm{\mathbf{r}^k}_2 \leq \epsilon^{pri}$ and $\norm{\mathbf{s}^k}_2 \leq \epsilon^{dual}$. These parameters can be tuned to get the desired convergence behavior. Should the convergence criterion not be met, TLDR will terminate after maximum number of iterations set by the user.

\subsection{Input Data}
Intended as a flexible method for reconstructing VDMs from RM data, TLDR makes minimal assumptions of the data provided to. As mentioned in Section~\ref{Sec:RM}, the continuum data must exist on an appropriate grid of times corresponding to the sample times of the spectral data and the delay times in the VDM to be reconstructed. In order to guarantee proper construction of the convolution operator TLDR applies a linear interpolation to the continuum data, but does not consider the uncertainty on the continuum. This is the only processing applied to continuum data by TLDR. Should continuum data be modelled at the appropriate times before being passed to TLDR, this linear interpolation will not affect the data in any way. Additionally, TLDR makes no assumptions on the spectral data provided to it. The option to model either the continuum data, or the spectra data is left to the user as data preparation. 

While TLDR does not require modelled data to be provided to it, modelling of both continuum data and spectral data prior is strongly recommended prior to reconstructing a VDM. As TLDR does not consider uncertainties on continuum data, modelling the continuum allows these uncertainties and large gaps in observations to be considered before any VDM is reconstructed, as well as providing a reliable method for interpolating the light curve. Gaussian Process modelling, previously adopted for Regularized Linear Inversion \citep{2015MNRAS.454..144S}, and Damped Random Walk models, as implemented in the JAVELIN code \citep{2011ApJ...735...80Z}, have successfully been used for modelling continuum light curves for RM. For a fast Gaussian Process modelling method easily applied to RM data see \cite{Foreman_Mackey_2017}. Similarly, modelling of spectral data with software such as PrepSPEC can isolate and remove constant spectral components, allowing TLDR to reconstruct a VDM based on only the broad line variations in the data. A recent discussion of the PrepSPEC software's spectral decomposition method can be found in \cite{2020arXiv200301448H}.

\subsection{Additional TLDR Notes}
In order to streamline the use of TLDR and ease frustrations that arise due to the sensitivity of the algorithm to the selection of the penalty parameters, three strategies for automatically adapting the penalty parameters were tested. Unfortunately, the simple strategy in \cite{Boyd:2011:DOS:2185815.2185816} $\S 3.4.1$, the more advanced residual balancing scheme described in \cite{wohlberg2017admm}, and the Adaptive Consensus ADMM (ACADMM) strategy of \cite{pmlr-v70-xu17c} resulted in instabilities in the algorithm. These instabilities arise out of the proximal projections used in TLDR, though they worked quite well when these terms were not used. As such, TLDR does not have a method of automatically selecting penalty parameters. This means that TLDR contains seven individual parameters that must be tuned by the user, which can be an arduous process for the user.

Tuning can be broken down into two categories, the penalty parameters ($\rho_X$, $\rho_P$, $\rho_N$, $\rho_T$) and regularization hyper-parameters ($\mu_{\ell_2}$, $\mu_{\ell_1}$, $\mu_{T}$). Increasing the magnitude of the penalty parameters reduces the number of iterations required before convergence. Varying the ratio between the various penalty parameters changes the relative importance of the different regularizers. Changing the regularization hyper-parameters increases the amount of smoothness or sparsity in the final VDM.

While tuning the various parameters for TLDR can be difficult, experience with the algorithm has revealed a general strategy. As a first step, set the penalty parameters to the same value and scale that value until TLDR converges in approximately 5000 iterations. Second, tune the regularization hyper-parameters. Beginning with each regularization parameter set to 1.0, tuning should be carried out one regularizer at a time with quadratic smoothing $\mu_{\ell_2}$ first, sparsity $\mu_{\ell_1}$ second, and TV $\mu_{T}$ last. Increase the value of the regularization parameter until the final reduced-$\chi^2$ reaches an inflection point and repeat for the next regularizer. Third, tune the relative ratio of the penalty parameters to move the final reduced-$\chi^2$ towards unity. Generally, this step has required increasing the penalty parameter for positivity more than anything else. It is likely that the user will need to iterate on the second and third steps to achieve a satisfactory result.

The TLDR algorithm is written in the Julia programming language \mbox{\citep{doi:10.1137/141000671}}. For ease and functionality, regularization variables exist as data structures within the algorithm's code. This allows the variables associated with each term to be stored and passed in the algorithm as a part of the regularization variable to which they belong.

\section{Testing}\label{Sec:Testing}
In order to test TLDR, synthetic emission line data  has been created using a number of model VDMs convolved with a continuum light curve. To generate  the synthetic emission line data the Johnson B band continuum light curve for Arp 151 from the Lick AGN Monitoring Project (LAMP) is adopted, the complete discussion of the photometric observing campaign and data reduction can be found in \cite{2009ApJS..185..156W}. Observations of Arp 151 took place between 2008 February 10 and May 16 and the resulting data were calibrated via differential photometry. The original continuum data has further been reduced by binning any observations made within 0.1 days.  The resulting light curve consists of 66 observations with a median sampling frequency of 1.02 days, mean sampling frequency of 1.5 days, mean flux of $\left<f\right>=0.93$ mJy with a standard deviation of 0.15 mJy, and flux ratio $f_{max}/f_{min} = 1.8$. The light curve has an excess variance of  $F_{var} = 0.16$ calculated as

\begin{equation}\label{eqn:excessvariance}
F_{var} = \frac{\sqrt{\sigma^{2} - \delta^{2}}}{\left< f \right>},
\end{equation}
 
 \noindent where $\sigma^2$ is the variance of the flux and $\delta^2$ is the mean square of the uncertainties of the flux. The continuum light curve is linearly interpolated onto a grid based on the date of spectroscopic observation and delay sample times. The dates used for the spectroscopic observations are those from the spectroscopic monitoring of Arp 151 from the LAMP campaign which range from 2008 March 25 to May 21 with a median sampling of 1.02 days and a mean sampling of 1.4 days, the full description of the spectroscopic observations and data reduction are provided in~\cite{2009ApJ...705..199B}. For all tests a delay sampling time of 1.0 days was used for the reconstruction. From the interpolated continuum light curve the convolution operator is created. Synthetic emission line data is then generated and Gaussian noise is added,

\begin{equation}\label{eqn:gensynth}
\mathbf{L}_{{synthetic}}=\mathbf{HX}_{{synthetic}}+\mathbf{n}.
\end{equation}

The continuum data used to generate all synthetic emission line data used for testing is shown in the top plot in Fig.~\ref{fig:RingRecovery_Curves} where the black circles represent the continuum observations of Arp 151 with error bars included and the black line represents the linear interpolation of those observations. By feeding the synthetic emission line data and the real continuum data into TLDR, the synthetic VDM is reconstructed. This provides a measure of the reconstructive efficacy of TLDR. Using the continuum data and sampling dates from actual observations of Arp 151 in the creation of the artificial emission line data creates a dataset with realistic cadences, but none of the following tests are intended to be compared to any reconstruction of the VDM for the Arp 151 dataset.

\subsection{Basic Tests}\label{subsec:Basic}
\begin{figure*}
\begin{center}
	\includegraphics[width=500px]{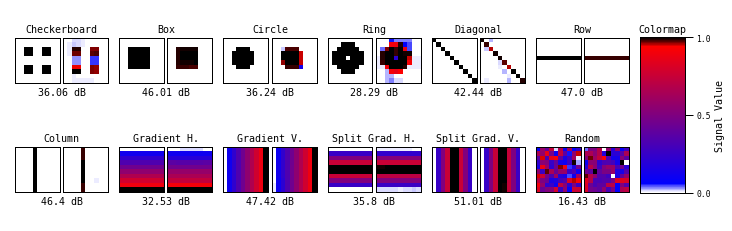}
    \caption{Reconstruction of 10x10 pixel test VDMs with signal values ranging from 0.0 to 1.0 and a signal to noise ratio of 50 in the synthetic emission lines. In each pair, the left plot shows the test VDM and the right plot shows the reconstructed VDM. Beneath each pair is the PSNR value of the reconstructed VDM in decibels. On the far right is the colormap used for each inset plot which emphasizes deviations from the extrema of the test VDMs making problem areas in the reconstructed VDMs stand out.}\label{fig:RecSwatches}
    \end{center}
\end{figure*}
In order to test the behavior of the TLDR algorithm a number of simple VDMs were used. Each test VDM consists of a $10 \times 10$ pixel matrix with signal values ranging between a minimum of 0.0 and maximum of 1.0. The test datasets were prepared with emission line errors yielding a SNR of 50.0. These test VDMs and their reconstructions shown in Fig.~\ref{fig:RecSwatches} where the original VDM appears on the left of each pair with the reconstruction on the right. On the far right is the colormap used for all of the test reconstructions, this colormap was specifically chosen to emphasize deviations from the expected values in the VDMs. Below each pair the peak signal to noise ratio (PSNR) of the reconstructed VDM is displayed providing a measure of the reconstruction fidelity. Where the PSNR is calculated in the usual way with
\begin{equation}\label{eqn:PSNR}
    PSNR = 20 \cdot \log_{10}\left( \frac{1}{\sqrt{MSE}}\right)
\end{equation}
and
\begin{equation}
MSE = \frac{1}{N}\sum_{i=1}^{N} \left( \mathbf{X}_i-\mathbf{X}_{true,i}\right)^2.
\end{equation}

The geometries were chosen to test the ability of TLDR to recover characteristics which may be encountered in real data. Clearly, the algorithm excels at reconstructions of smooth gradients and places where test VDMs have continuous values across their cross-sections. This is unsurprising as the majority of the regularizers used in TLDR promote smoothness. The algorithm has some difficulty when the VDM being reconstructed contains abrupt signal change as in the Checkerboard, Box, Circle, Ring, and Diagonal test VDMs. Unsurprisingly, TLDR struggles most when reconstructing a VDM consisting purely of random noise.

\begin{figure}
	\begin{center}
	\includegraphics[width=190px]{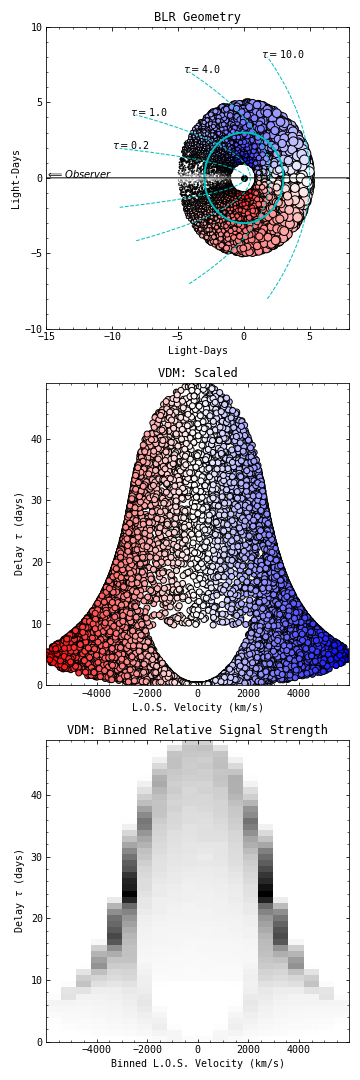}

 \caption{Top: Overview of the simulated flat Keplerian disk showing 10,000 particles in the same two-dimensonal isodelay surface as Fig.~\ref{fig:isodelay}. The size of the markers shows relative delay and color indicates red/blue-shift due to rotational velocity. Middle: Projected VDM of the particles in the simulation, again color indicates red/blue-shift. Bottom: Binned VDM of the simulated data. } \label{fig:simdisk}

\end{center}
\end{figure}

\subsection{Synthetic VDM of Keplerian disk}\label{subsec:SyVDM}
In order to test TLDR with a more complicated VDM than those used in Section~\ref{subsec:Basic}, a synthetic VDM was created using a simple flat Keplerian disk toy model of the BLR. This synthetic VDM was generated from an array of 1 million randomly situated points arranged in a flat disk about a central point corresponding to the location of a SMBH with mass $7 \times 10^6$ $M_{\odot}$ with an inner radius of 1.0 and an outer radius of 5.0 light days.

As demonstrated in other work on RM \citep{1991ApJ...379..586W,2004AN....325..248P},  the line of sight velocities $\mathbf{v}$ and delay times $\boldsymbol{\tau}$ for particles in an edge-on circular disk $\mathrm{i=90}$ are easily found from the geometry and the orbital velocity $V$ at radius $r$, where the angle $\phi$ is the azimuth angle in the plane of the disk starting from zero at the line of sight to the observer.

\begin{equation}\label{eqn:VLOS}
\mathbf{v}\left(r,\phi\right) = V\left(r\right)\sin\left(\phi\right)
\end{equation}
and
\begin{equation}\label{eqn:Delay}
\boldsymbol{\tau}\left(r,\phi\right) = \frac{r}{c}\left(1-\cos\left(\phi\right)\right).
\end{equation}

\noindent After calculating the line of sight velocities and delays for all of the points in the simulated disk, the points are projected onto the VDM seen in Fig.~\ref{fig:simdisk} middle panel. The VDM is then binned along both delay time and velocity (Fig.~\ref{fig:simdisk} lower panel) with the delay times scaled to a maximum of 50.0 light-days. The signal value of each pixel is normalized and scaled by counting the number of points that lie within a given pixel, normalizing to the maximum number of points found in a single pixel, and scaling the normalized array to the desired maximum value. Finally, each pixel is scaled by the ratio of its delay to the maximum delay to enhance the structure that appears in the VDM. In this case, the VDM was binned with 50 pixels of 1.0 light-day along the temporal axis and 20 pixels on the spectral axis which happen to be 597 km/s wide, with a maximum signal value of 0.1.

\begin{figure}
	\begin{center}
	\includegraphics[width=250px]{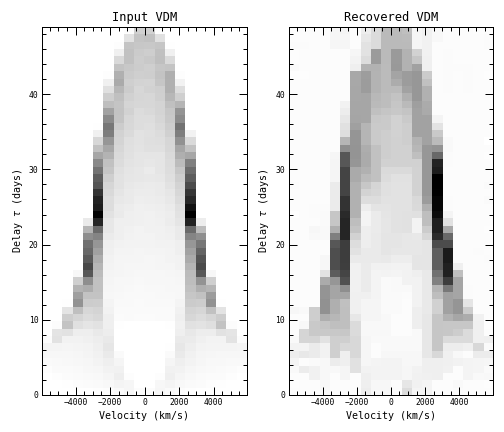}

 \caption{Synthetic VDM of a Keplerian disk (left) and the recovered VDM (right). Recovered VDM shows much of the detail of the synthetic VDM, with some artifacting visible in the recovered VDM.}\label{fig:RingRecovery}

\end{center}
\end{figure}

This toy model is over-simplified in that the binned flux does not account for the relative flux difference in different delays corresponding to the light-travel differences. Furthermore, the scaling and binning used for the reconstruction change the physicality of the model. Similarly, the use of a flat Keplerian disk is simplistic. The goal in generating this synthetic VDM is to provide a test case showing the algorithm's capability for recovering complicated signals that mimic those recovered by other methods. This synthetic VDM is not intended as a realistic model of the BLR. Furthermore, despite borrowing much from the Arp 151 dataset to create this synthetic line emission data, this toy model is not intended to provide any physical comparison to any reconstruction of a VDM for the Arp 151 dataset.

The synthetic VDM was used to create synthetic spectra according to equation~\ref{eqn:gensynth} with a fractional uncertainty of $1.5\%$ matching the mean fractional error from the real Arp 151 dataset\citep{2009ApJ...705..199B} and then reconstructed using TLDR. The reconstructed synthetic VDM generated from the Keplerian disk model and Arp 151 continuum is shown alongside the original synthetic VDM in Fig.~\ref{fig:RingRecovery}. Comparing the reconstructed Keplerian disk VDM to the original  shows that the algorithm is providing an excellent reconstruction. There is some noisy distortion apparent at longer delay times and some loss of fine detail near the edges of the VDM, but with a PSNR of 45.0 dB the details of the VDM are well reconstructed.

\begin{figure}
	\begin{center}
	\includegraphics[width=250px]{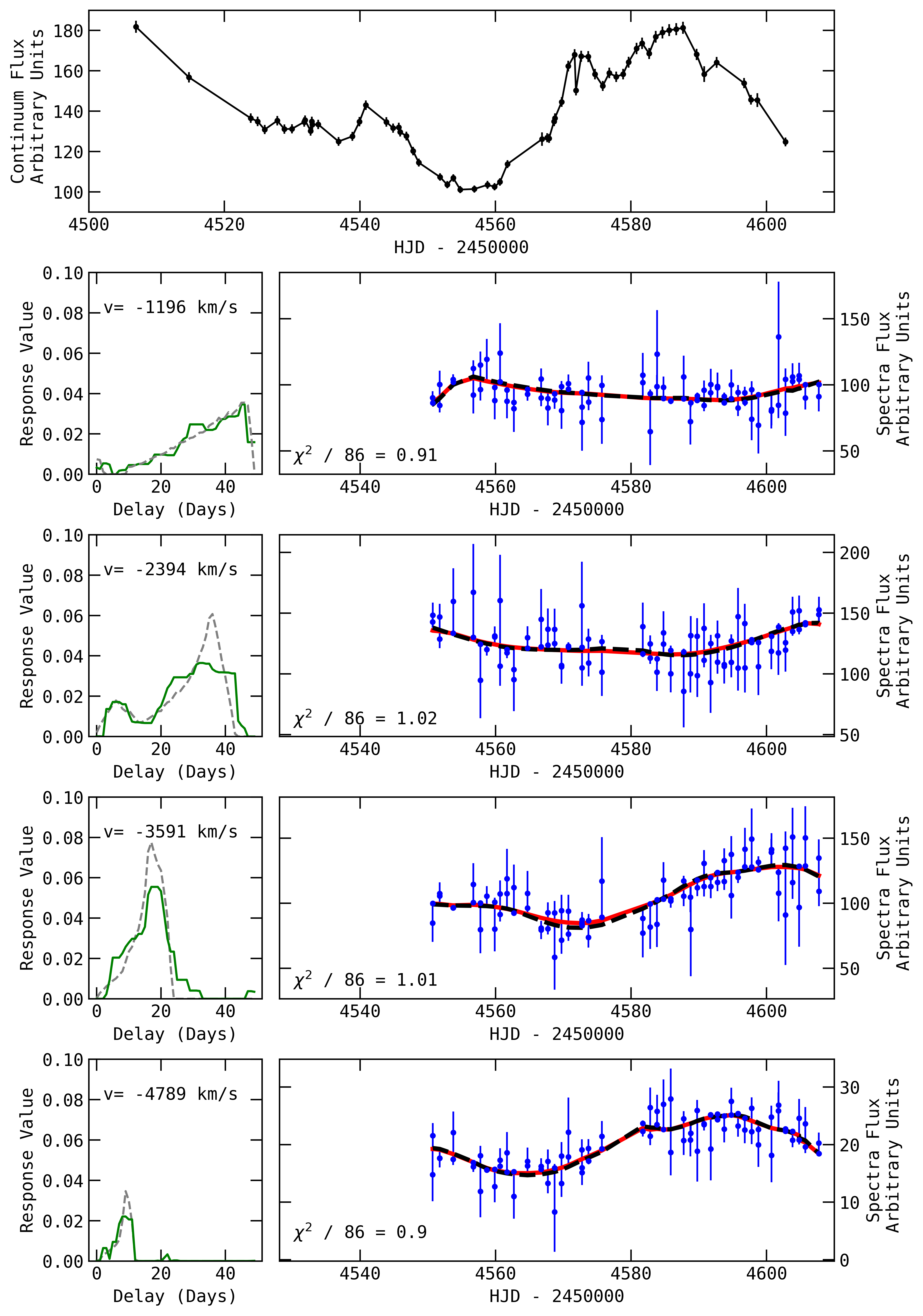}

 \caption{Sample light curves from the reconstruction of the Keplerian disk. Top: Continuum light curve where the black circles show the original sample data with error bar and the black solid line shows the linear interpolation of the original data. Bottom four rows: On the left, response function where the grey dashed line is the true function and the green solid line is the recovered function. On the right: Noiseless spectral light curve shown by the black dashed line, the noisy spectral light curve shown as the blue circles and the recovered spectral light curve as the red solid line. The lines shown represent 20\% (-1196 km/s), 40\% (-2394 km/s), 60\% (-3591 km/s), and 80\% (-4789 km/s) the maximum line of sight rotational velocity of the Keplerian disk model used to generate these emission line data.}\label{fig:RingRecovery_Curves}

\end{center}
\end{figure}

The reconstruction can be further evaluated by looking at a sample of the light curves from the reconstruction shown in Fig.~\ref{fig:RingRecovery_Curves}. The top plot is simply the continuum light curve used to generate the synthetic emission line data (i.e. the original Arp 151 continuum data) shown by black circles with error bars slightly larger than the markers and the interpolated data shown as the black solid line. The bottom four rows show the spectral light curves from the data at 20\% (-1196 km/s), 40\% (-2394 km/s), 60\% (-3591 km/s), and 80\% (-4789 km/s) of the maximum line of sight rotational velocity of the model. Each shows the response function on the left with the true response as the grey dashed line and the reconstructed response as the green solid line. On the right each shows the spectral light curve with the noiseless light curve shown as a black dashed line, the noisy spectral light curve input as the blue circles, and the recovered spectral light curve as the red solid line. Each also contains the line's reduced $\chi^2$ value. While in each case, the recovered spectral line closely matches the noiseless true spectral line, they all show deviation from the true response function in the recovered response functions. The second row showing the curves for the -1196 km/s cross-section shows a very good recovery of the response function and spectral light curve. The bottom three rows, showing the -2394 km/s, -3591 km/s, and -4789 km/s lines contain over-smoothed responses where the signals assume the general shape expected but are suppressed in the regions of maximum response value. It would likely be possible to achieve a better fit across all lines by using a different set of regularization hyper-parameters for each line, but this would further complicate what can already be an arduous parameter selection process, a trade off which must be considered for any image reconstruction process.

\section{Preliminary Reconstruction of Arp 151 \texorpdfstring{H$\beta$}{Hbeta}} \label{Sec:TestingArp151}
\begin{figure}
	\begin{center}
	\includegraphics[width=250px]{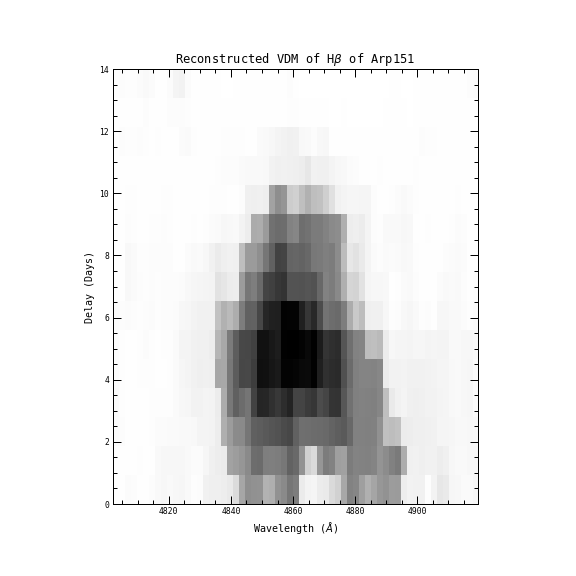}

 \caption{Reconstructed TDF of the $H\beta$ feature of Arp 151 from the LAMP RM project.}\label{fig:RDVDM}

\end{center}
\end{figure}

As a final indication of the utility of TLDR, an initial reconstruction of the VDM for the $H\beta$ feature of Arp 151 from LAMP \citep{2009ApJ...705..199B} is provided. Using the continuum subtracted model spectra and the Johnson B continuum data used in for the Maximum Entropy Method reconstruction presented in \cite{2010ApJ...720L..46B}, an initial VDM was calculated in accordance with Section~\ref{SubSecInitialVDM} with a smoothing hyper-parameter of $2.0\times10^6$. The VDM was set up with the pixel spectral width matching the $2.0 \textup{\AA}$ steps of the original spectral observations and a temporal width of 1.0 days matching the global average sampling frequency of the continuum and spetral observations. For this reconstruction, TLDR was run for $3.0\times10^4$ iterations with the parameters listed in Table~\ref{tab:RecParams}, selected from a grid of possible regularization hyper-parameters by proximity to $\chi^2=1.0$. The reconstruction terminated with a reduced $\chi^2$ of 1.02, yielding the VDM shown in Fig.~\ref{fig:RDVDM}. Visual comparison between the VDM reconstructed with TLDR and those recovered by the Maximum Entropy Method \citep{2010ApJ...720L..46B} and Regularized Linear Inversion \citep{2015MNRAS.454..144S} largely agree. In each case, the overall shape of the response in the VDM appears to match with asymmetries appearing  in similar locations. However, to make any substantive comparison between results a thorough comparative analysis would be necessary. 

\begin{table*}
\caption{Reconstruction Parameters used in the reconstruction of the $H\beta$ feature of Arp 151.}
\label{tab:RecParams}
 \begin{tabular}{c c c} 
 
\hline\hline 
Regularizer & Hyper-Parameter $\mu$ & Penalty-Parameter $\rho$   \\ [0.5ex] 
\hline 
Smoothing $\bf{X}$& $2.0\times10^6$ & 1.0 \\
Positivity $\bf{P}$ & N/A & 100.0 \\
Sparsity $\bf{N}$ & 10.0 & 1.2 \\
Total Variation $\bf{T}$ & 1.75 & 1.2 \\
\\
\hline
\end{tabular}

\end{table*}

\section{Conclusion}
The TLDR algorithm provides an additional tool for RM, that can be used in concert with existing tools or independently, to reconstruct VDMs from RM data. The regularization scheme implemented within TLDR provides a flexible platform for reconstruction of VDMs. With numerous recent and ongoing RM campaigns \citep{2015ApJ...806..128D,2016ApJ...818...30S,2017AAS...22941404U}, interest in using RM to study AGNs is high. With high interest, and a growing number of datasets, TLDR should be able to make a sizable contribution to the RM field.

Future work on TLDR will include applying the algorithm to a number of the well studied datasets from the Lick AGN Monitoring Project (LAMP) \citep{2009ApJ...705..199B}. The data contained in the LAMP data release has been studied by all of the reverberation mapping algorithms currently in use, including the Maximum Entropy Method \citep{2010ApJ...720L..46B}, Regularized Linear Inversion \citep{2015MNRAS.454..144S}, as well as direct modeling \citep{2014MNRAS.445.3055P}. This will provide a way to directly evaluate the performance of TLDR and compare the different methods currently available.

\section{Acknowledgements}

FB and MA acknowledge support from the National Science Foundation through AAG grants AST-1210972, AST-1616483 and AST-1814777. MCB gratefully acknowledges support from the National Science Foundation through CAREER grant AST-1253702 and AAG grant AST-2009230 to Georgia State University. We thank the referee for their thoughtful comments and thorough review.

\section{Data Availability}
The TLDR software repository is publicly available at https://github.com/matdander/TLDR. The data underlying this article were accessed from the Lick AGN Monitoring Project\citep{2017AAS...22941404U} at https://www.physics.uci.edu/~barth/lamp.html. 

\newpage





\bibliographystyle{mnras}
\bibliography{references}




\bsp	
\label{lastpage}
M.D. Anderson\end{document}